\documentclass[]{spie}

\usepackage{amsmath,amsfonts,amssymb}
\usepackage{graphicx}
\usepackage[colorlinks=true, allcolors=blue]{hyperref}
\usepackage{multirow,bigdelim}

\def\CN2{\mbox{$C_N^2 \ $}}

\def\CT2{\mbox{$C_T^2 \ $}}

\def\sigmal2{\mbox{$\sigma ^{2}_{I} \ $}}

\title{Forecasts of the atmospherical parameters close to the ground at the LBT site in the context of the ALTA project} 

\author[a]{Alessio Turchi}
\author[a]{Elena Masciadri}
\author[a]{Luca Fini}
\affil[a]{INAF - Osservatorio Astrofisico di Arcetri, L.go E. Fermi 5, 50125  Florence, Italy}

\authorinfo{Correspondence to Alessio Turchi: aturchi@arcetri.astro.it}

\begin{document} 
\maketitle 

\begin{abstract}
In this paper we study the abilities of an atmospherical mesoscale model in forecasting the classical atmospherical parameters relevant for astronomical applications at the surface layer (wind speed, wind direction, temperature, relative humidity) on the Large Binocular Telescope (LBT) site - Mount Graham, Arizona. The study is carried out in the framework of the ALTA project aiming at implementing an automated system for the forecasts of atmospherical parameters (Meso-Nh code) and the optical turbulence (Astro-Meso-Nh code) for the service-mode operation of the LBT. The final goal of such an operational tool is to provide predictions with high time frequency of atmospheric and optical parameters for an optimized planning of the telescope operation (dome thermalization, wind-dependent dome orientation, observation planning based on predicted seeing, adaptive optics optimization, etc...). Numerical simulations are carried out with the Meso-Nh and Astro-Meso-Nh codes, which were proven to give excellent results in previous studies focused on the two ESO sites of Cerro Paranal and Cerro Armazones (MOSE Project). In this paper we will focus our attention on the comparison of atmospherical parameters forescasted by the model close to the ground with measurements taken by the observatory instrumentations and stored in the LBT telemetry in order to validate the numerical predictions. As previously done for Cerro Paranal (Lascaux et al., 2015), we will also present an analysis of the model performances based on the method of the contingency tables, that allows us to provide complementary key information with the respect to the bias and RMSE (systematic and statistical errors), such as the percentage of correct detection and the probability to obtain a correct detection inside a defined interval of values. 
\end{abstract}



\section{INTRODUCTION}
\label{sec:intro} 
This paper is part of a general validation study on the forecasts of meteorological parameters and optical turbulence (OT) at the Large Binocular Telescope (LBT) site of Mount Graham (Arizona), performed in the context of Advanced LBT Turbulence and Atmosphere (ALTA\footnote{\url{http://alta.arcetri.astro.it}}) Center project. The ALTA project aims to implement and automate forecast system for LBT using a mesoscale hydrodynamic meteorological model, either for classical meteorological parameters (wind speed and direction, temperature, relative humidity) which are relevant for ground-based astronomy, and astroclimatic parameters ($C_N^2$ profiles, seeing $\epsilon$, isoplanatic angle $\theta_0$, wavefront coherence time $\tau_0$) which are relevant for adaptive optics applications (AO). The final outcome of ALTA project will be the deployment of an operational tool to provide predictions with high time frequency and spatial resolution in order to support and optimize the telescope operation, such as dome thermalization, wind-dependent dome orientation, observation planning  based on seeing and other OT parameters condition and optimization of AO systems operation. The first commissioning for the atmospheric parameters is set in June 2016, while the second commissioning for the OT parameters is on December 2016.\\
In order to provide the above results we use MESO-Nh (Lafore et. al. 1998 [\cite{lafore98}]) model developed by the Laboratoire d'Aerologie, CNRM and Meteo France, together with the Astro-Meso-Nh  module (Masciadri et. al. 1999 [\cite{masciadri99a}]) which is used to provide forecast of OT parameters. The forecasts are relative to the night time frame, when the LBT telescope is operative.\\
The MESO-Nh model performances for the OT forecasts have already been tested in previous studies performed on major telescope installations, such as Roque de los Muchachos[\cite{masciadri2001a}], San Pedro Martir[\cite{masciadri2004}], Cerro Paranal[\cite{masciadri99b}] and also at very low latitudes such as Antarctica[\cite{lascaux2009,lascaux2010}].  In terms of performances and reliability the critical milestones have been: the proposition of a technic for the model calibration (Masciadri and Jabouille, 2001[\cite{masciadri2001}]) and the validation of such a technique with a sample of 10 nights (a large sample for that epoch) (Masciadri et al., 2004[\cite{masciadri2004}]). A few years ago the model has been validated at Mount Graham (Hagelin et. al. 2011 [\cite{hagelin2011}]) using measurements of a Generlized SCIDAR related to the more extended site testing campaign performed at Mt. Graham (43 nights) (Masciadri et. al 2010 [\cite{masciadri2010}]). Finally the model forecasts for atmospheric and astroclimatic parameters were part of a large validation campaign conducted within the MOSE project, commissioned by the European Souther Observatory (ESO) in order to prove the feasibility of an automated forecast system for their installations in Cerro Paranal and Armazones (VLT and E-ELT respectively) (Masciadri et. al. 2013 [\cite{masciadri2013}], Lascaux et. al. 2013-2015[\cite{lascaux2013,lascaux2015}]).\\
In this paper we will show the preliminary results of an ongoing validation study, in view of the first commissioning of June 2016. In this context, we will limit ourselves to study the performance of Meso-Nh model in forecasting the standard atmospheric parameters near the ground layer, using as a reference the LBT weather stations placed on the telescope dome. The study is performed on a limited sample of nights, which was used to test the model stability and tune the settings in order to obtain the most efficient configuration for an operational tool.\\
A more detailed study, performed on a larger sample of nights, will be published in a forthcoming paper.\\
In section \ref{sec:obs} we will describe the LBT site and the measurements sets that were use for the present validation study. In section \ref{sec:mod_conf} we will describe the model configuration, including the numerical setup used for this study. In section \ref{sec:res} we will present the results of the validation in terms of statistical operators and contingency tables. Finally in section \ref{sec:concl} we will draw the conclusions.\\
\section{MEASUREMENTS AND TEST SITE CONFIGURATION}
\label{sec:obs} 
\begin{figure} [ht]
\begin{center}
\begin{tabular}{cc} 
\includegraphics[height=5cm]{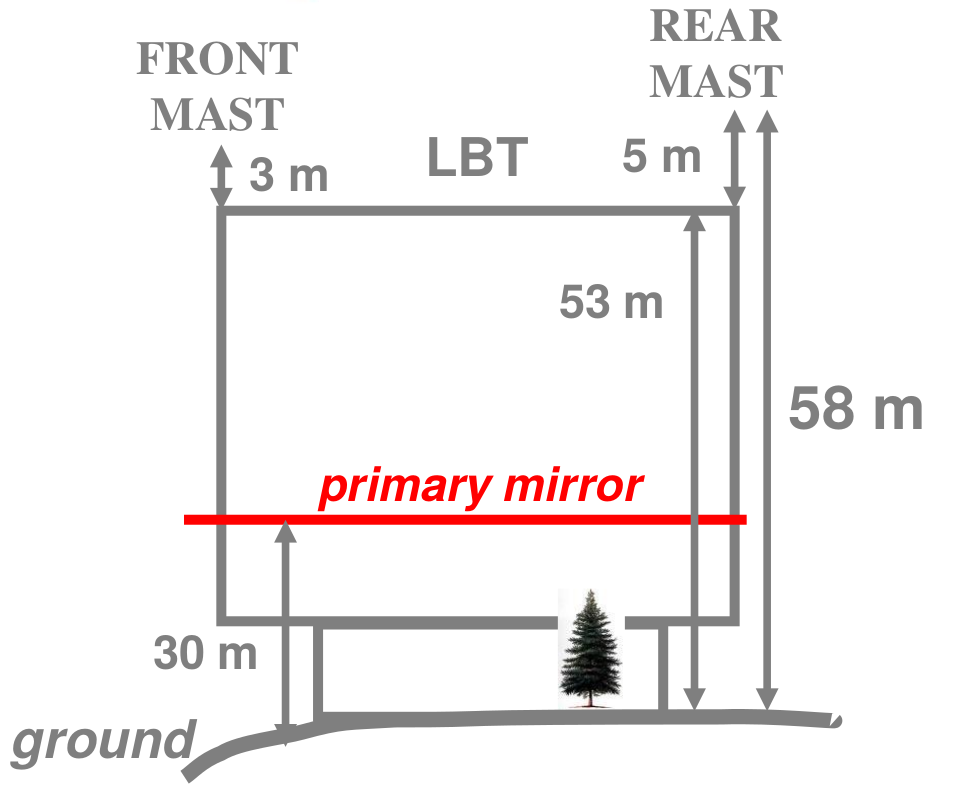} & \includegraphics[height=5cm]{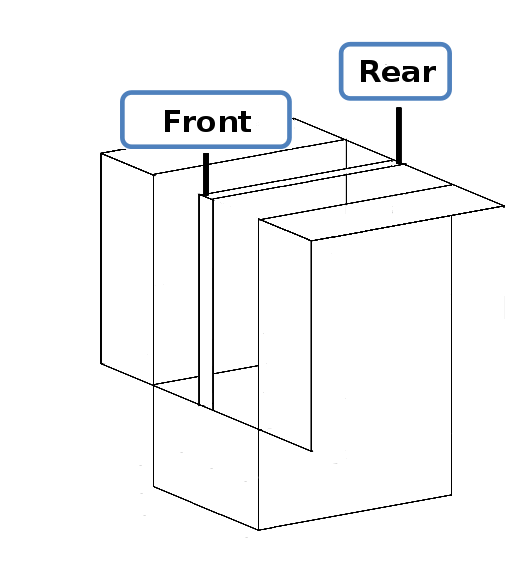}
\end{tabular}
\end{center}
\caption[example] 
{ \label{fig:lbtmast} 
Large Binocular Telescope dome scheme with reported weather mast locations.}
\end{figure}

Measurements of atmospheric parameters close to the ground are taken from the telemetry data provided by LBTO, streamed directly from the instruments placed at the telescope site. LBT has two weather stations, each operated independently, both placed on a mast above the telescope dome. As shown in Fig.\ref{fig:lbtmast} the first station, labeled ``FRONT'', is placed on the front side of the roof, with respect to the mirrors orientation, right between the two dome apertures, on a mast which is 3~m above the roof and 56~m above ground level. The FRONT station is provided with an anemometer measuring wind speed and direction. The second station, labeled ``REAR'', is placed on the rear side of the roof, in a centered position. This station is provided with an anemometer at 5~m above the roof and 58~m above the ground, measuring both wind speed and direction, and with a set of sensors measuring temperature, relative humidity and pressure, placed at 2.5~m above the roof and 55.5~m above the ground level on the same mast. Data from the weather stations are sent to telemetry streams and stored in HDF5 file format on LBTO servers. Stored data is sampled without a precise time step, however, approximately, telemetry provides one sample per second. In our analysis we reduced the time step of the measurements by taking an average over each minute of data in order to have a precise sampling time of one minute.\\


In this preliminary validation study we selected a 22 nights sample, reported in Table \ref{tab:dates} distributed between October 2015 and December 2015, based on the criterion of having a whole full night worth of data with no registered telemetry fail.\\
\begin{table}[ht]
\caption{Selected nights (UT) for preliminary model validation.} 
\label{tab:dates}
\begin{center}       
\begin{tabular}{|c|c|c|c|c|c|c|} 
\hline
\rule[-1ex]{0pt}{3.5ex}  2015/10/24 & 2015/10/25 & 2015/10/26 & 2015/10/28 & 2015/10/29 & 2015/11/03 & 2015/11/08\\
\hline
\rule[-1ex]{0pt}{3.5ex}  2015/11/09 & 2015/11/10 & 2015/11/14 & 2015/11/24 & 2015/11/25 & 2015/11/28 & 2015/11/30\\
\hline
\rule[-1ex]{0pt}{3.5ex}  2015/12/05 & 2015/12/10 & 2015/12/11 & 2015/12/14 & 2015/12/16 & 2015/12/19 & 2015/12/20\\
\hline
\rule[-1ex]{0pt}{3.5ex}  2015/12/21 & & & & & &  \\
\hline
\end{tabular}
\end{center}
\end{table}

The above sample size was chosen in order to perform a large preliminary test campaing on all the possible model configurations. In a forthcoming study the sample size will be increased to $\sim 140$ nights uniformly distributed between 2014 and 2015 in order to provide a more reliable statistical validation.

\section{MODEL CONFIGURATION}
\label{sec:mod_conf} 

Meso-Nh\footnote{\url{http://mesonh.aero.obs-mip.fr/mesonh/}} is an atmospherical mesoscale model that simulates the time evolution of weather parameters in a three-dimensional volume over a finite geographical area with a forward in time (FIT) numerical scheme. The coordinate system is based on mercator projection, since we are at low latitudes, while the vertical levels use the Gal-Chen and Sommerville coordinate systems (Gal-Chen et. al. 1975 [\cite{chen}]). 
The model is based on anelastic formulation of hydrodynamic equations, in order to filter acoustic waves. Simulations are made using a one-dimensional mixing length proposed by Bougeault and Lacarrere (Bougeault et. al. 1989 [\cite{Bougeault89}]) with a one-dimensional 1.5 closure scheme (Cuxart 2000 [\cite{Cuxart00}]). The exchange between surface and atmosphere is computed with the Interaction Soil Biosphere Atmosphere - ISBA scheme (Noilhan et. al. 1989 [\cite{Noilhan89}]).\\
We initialise the model simulations using initialisation data provided by the European Center for Medium Weather Forecasts (ECMWF), calculated with their General Circulation Model extend on the whole globe, with an horizontal resolution of 16~km\footnote{Since March 2016 resolution increased to roughly 9~km.}. Simulation are initialised for each night at 00:00 UT and forced every 6 hours with new data from ECMWF. Simulations run for 15 hours up to 15:00 UT in order to completely cover the night time period of Mount Graham.\\
The LBT site is located at Mt. Graham ($[32.70131, -109.88906]$) at an height of 3221~m above sea level. In order to define the computation grid for the simulations, we use a grid-nesting technique (Stein et. al. 2000 [\cite{Stein00}]). This consists of using different imbricated domains, described in Table \ref{tab:resol}, with digital elevation model (DEM, i.e. orography) extended on smaller and smaller surfaces having a progressively higher horizontal resolution. In this way, using the same vertical grid resolution, we can achieve a higher horizontal resolution on a sufficiently small scale around the summit to provide the best possible prediction at the specific site. Each domain is centered on LBT coordinates.\\

\begin{table}[h]
\caption{Horizontal resolution of each Meso-Nh imbricated domain.} 
\label{tab:resol}
\begin{center}       
\begin{tabular}{cccc} 
\hline
\rule[-1ex]{0pt}{3.5ex}  Domain & $\Delta$X (km) & Grid points & Domain size (km) \\
\hline
\rule[-1ex]{0pt}{3.5ex}  Domain 1 & 10 & 80x80 & 800x800 \\
\rule[-1ex]{0pt}{3.5ex}  Domain 2 & 2.5 & 64x64 & 160x160 \\
\rule[-1ex]{0pt}{3.5ex}  Domain 3 & 0.5 & 120x120 & 60x60 \\
\rule[-1ex]{0pt}{3.5ex}  Domain 4 & 0.1 & 100x100 & 10x10 \\
\hline
\end{tabular}
\end{center}
\end{table}
The DEM used for domains 1 and 2 is the GTOPO\footnote{\url{https://lta.cr.usgs.gov/GTOPO30}}, with an intrinsic resolution of 1~km. In domains 3 and 4 we use the SRTM90\footnote{\url{http://www.cgiar-csi.org/data/srtm-90m-digital-elevation-database-v4-1}} (Jarvis et. al. 2008 [\cite{srtm}]), with an intrinsic resolution of approximately 90~m (3 arcsec).\\
In our configuration the grid-nesting allows a 2-way interaction between the interface of each domain with the containing (larger) one (the so called father-son domains). Under these conditions the atmospheric flow in the inner domains is in a constant thermodynamic equilibrium with the outer domain's flow, allowing for the propagation of gravity waves through the whole area mapped by the simulation independently on the specific domain.\\
We use 54 physical vertical levels on each domain, with the first grid point equal to 20~m above ground level (a.g.l.), and a logarithmic stretching of 20\% up to 3.5~km a.g.l. From this point onward the model uses an almost constant vertical grid size of $\sim$ 600~m up to 23,57 km, which is the top level of our domain. The grid mesh deforms uniformly to adapt to the orography, so the actual size of the vertical levels can stretch in order to accommodate for the different ground level at each horizontal grid point. As indicated in section \ref{sec:obs}, LBT weather stations are positioned $\sim $ 55-58~m above ground. At the LBT coordinates, this corresponds to the third physical Meso-Nh level spanning the interval [38-62]~m.\\
The model has been managed so to access to the temporal evolution of the surface parameters calculated on the summit (LBT location) with a temporal frequency equal to the time step of the innermost domain. In our model configuration this is of the order of a few seconds. We used 3~s for those parameters that we analyse with a three domains configuration (temperature, relative humidity, wind direction) and 1~s for those parameters (wind speed) that we analyse with four domains. 
As proved by Lascaux et al. 2013[\cite{lascaux2013}], an horizontal resolution of 100~m is necessary to well reconstruct the wind speed close to the ground particularly when the wind speed is strong.\\

\section{MODEL VALIDATION}
\label{sec:res}
To compare measurements and simulation outputs, similarly to what has already been done in previous studies on Cerro Paranal and Cerro Armazones (Lascaux et. al. 2013-2015 [\cite{lascaux2013,lascaux2015}]), which were approved by ESO staff, we decided to perform a moving average over a 1-hour time window, from 30 minutes before to 30 minute after. This operation allows for the filtering of fast frequencies and consent us to estimate the performance over the slower-moving trends, which are of interest for the telescope operation and planning. Data were then resampled over 20-minutes intervals. The described operation was done on either observed values and simulation outputs and is preliminary to any statistical analysis.\\
The ratio between the above described operation is that astronomers are interested in identifying the trend of a parameter over a time sampling of at least 20 minutes, which is the least possible time to switch an instrument or change the telescope planning during the night.\\
Finally, we removed from the sample of data-set those data which are non-relevant for the telescope operation. We selected only values contained between sunset an sunrise hours, computed with ephemerid tables for each date. On average we start the data selection at $00$:$30$ UT and end at $14$:$00$ UT, for a total of $N=911$ samples.\\

In order to quantify the model reliability in reconstructing the weather parameters we used two different statistical approaches. First we analysed the classical BIAS, RMSE and $\sigma$ statistical operators, defined as:
\begin{equation}
BIAS = \sum\limits_{i = 1}^N {\frac{{(Y_i  - X_i )^{} }}
{N}} 
\label{eq1}
\end{equation}
\begin{equation}RMSE = \sqrt {\sum\limits_{i = 1}^N {\frac{{(Y_i  - X_i )^2 }}
{N}} } 
\label{eq2}
\end{equation}
where $X_{i}$ are the individual observations and $Y_{i}$ the individual simulations calculated at the same time index $i$, with $1\geq i\leq N$, $N$ being the total sample size.\\
From the above quantities we deduce the bias-corected RMSE ($\sigma$):
\begin{equation}\sigma = \sqrt {RMSE^2 - BIAS^2}
\label{eqr}
\end{equation}
The previously defined indicators provide us informations on the statistical and systematic errors of the model.\\

We then proceeded to perform a different analysis based on contingency tables, similar to what was already done for VLT on Cerro Paranal by Lascaux et al., 2015[\cite{lascaux2015}], which are very useful to provide complementary informations the reliability of the model in a realistic use-case scenario. A contingency table is a method to analyse the relationship between categorical variables. It consists in distributing observed and simulated values obtained respectively by a instrument and a model in tables delimited by intervals of values (see an example in Fig.\ref{tab:tertemp}). Starting from these contingency tables it is possible to evaluate the score of success of the model (and more generally the model performances) using different statistical operators. We refer to Lascaux at al. 2015 [\cite{lascaux2015}] for a detailed definition and description of contingency tables (2$\times$2, 3$\times$3 and 4$\times$4). For these atmospherical parameters it is reasonable to use 3$\times$3 tables for temperature, wind speed and relative humidity and a 4$\times$4 table for the wind direction. Quadrants are defined as follows: North=$[315^\circ, 45^\circ]$, East=$[45^\circ,135^\circ]$, South=$[135^\circ,225^\circ]$, West=$[225^\circ,315^\circ]$. We quantify the model performances using the following operators: the percentage of correct detection (PC), the probability of detection (POD) of a parameter in a particular range of values and the extremely bad detection (EBD) (see definitions of PC, POD and EBD in Lascaux et al. 2015 - Eq.9, 10,11,12 and 13).\\

In the case of a perfect forecast we would have PC=PODs=100\% and EBD=0, while in the worst case (random prediction) we would have PC=PODs=33\% and EBD=22.5\%. A number of others statistical parameters can be deduced from a contingency table. An extended literature exists in this respect. We will limit the current study to the analysis of the parameters mentioned above that already provide a sufficiently accurate description for this preliminary analysis.\\
As we said, in the case of temperature, relative humidity and wind speed, a 3$times$3 contingency table was used. For this preliminary study, we decided to use as threshold of the contingey tables categories the tertiles of the cumulative distribution of each parameter, computed over the 2014 and 2015 winters (October-March), for which we have abundance of data from the LBT telemetry system (Table \ref{tab:tertemp}). Also or sample of data test belong to this period of the year.

\begin{table}[ht]
\caption{Climatological tertiles for temperature at Mount Graham. Left column: $33\%$, central column: median, right column: $66\%$. Values are computed over the 2014-2015 winter distribution (October-March) of observations from LBT telemetry.} 
\label{tab:tertemp}
\begin{center}
\begin{tabular}{c|ccc}
\hline
 Mount Graham & $33\%$ & median $(50\%)$   &  $66\%$ \\
\hline
Temperature $(^\circ C)$ & -1.2 & 0.6 & 2.5\\
Relative humidity $(\%)$   & 28.4 & 46.6 & 71.7\\
Wind speed (m/s)    & 5.9 & 7.6 & 9.7\\
\hline
\end{tabular}
\end{center}
\end{table}

\subsection{Temperature}

\begin{figure}[ht]
\begin{center}
\begin{tabular}{cc}
\includegraphics[height=5cm]{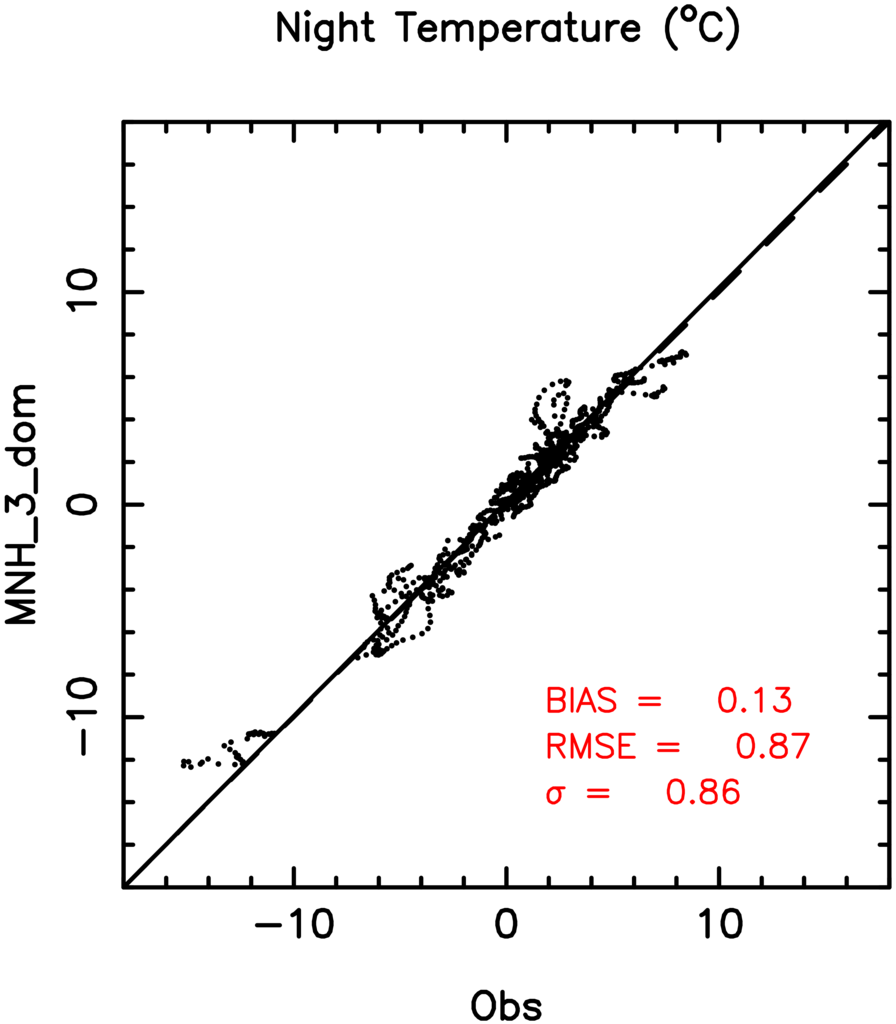} & \includegraphics[height=5cm]{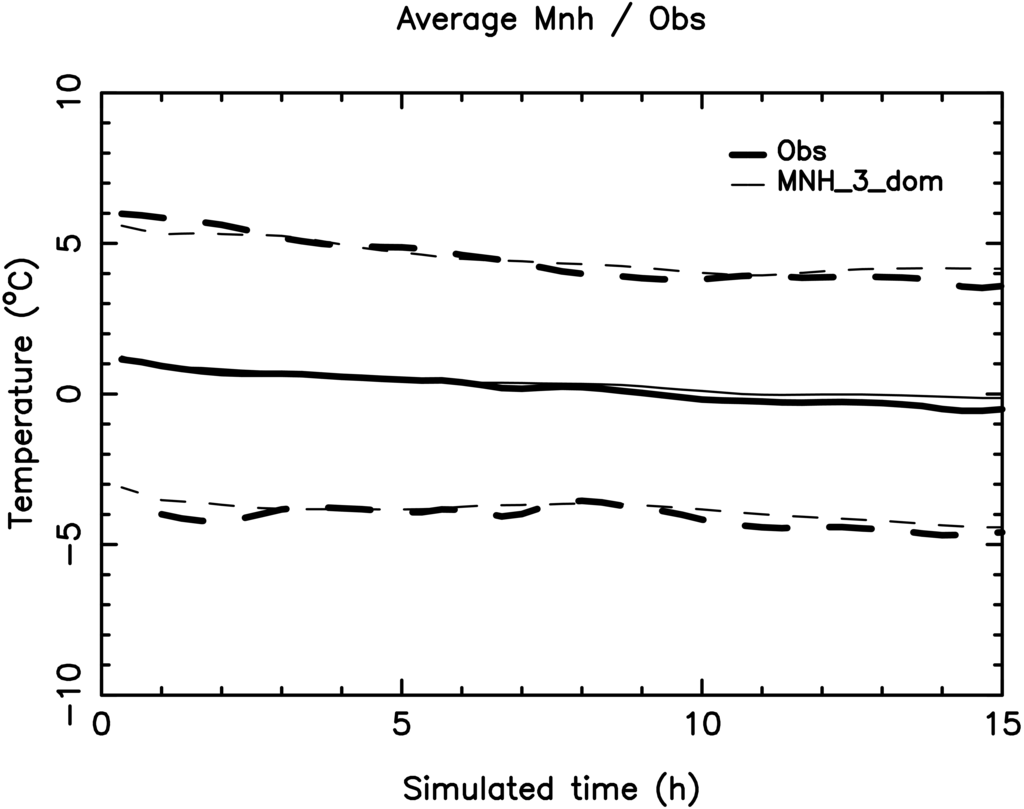} \\
\end{tabular}
\end{center}
\caption[example]{\label{fig:treg}Left: scatter plot for temperature, comparing model outputs (MNH) and measurements (OBS). The full black line is the regression line passing by the origin, while the dashed line represent the reference diagonal line for unbiased results.\\
Right: time evolution of the average of both model outputs (thin line) and measurements (bold line), from 00:00UT to 15:00UT.} 
\end{figure}

We first report the results obtained for the temperature on the scatter plot in Fig.\ref{fig:treg} (left side) calculated on the whole sample of 22 nights. For this parameter we used the model configuraiton at three domains with the horizontal resolution of the innermost domain equal to 500~m, since our tests showed us that this setup was sufficiently accuracy to obtain reliable predictions. Measurements show a distribution of temperatures, mostly between -10 $^\circ C$ and 10 $^\circ C$, with few data below -10 $^\circ C$, which are typical on Mount Graham summit during winter time. In Fig.\ref{fig:treg} (right side) we also report the time evolution of the average values of both Meso-Nh forecasts and measurements, each computed as the ensemble average of the time evolutions of the temperature over each sample defined in Table \ref{tab:dates}. The figure shows no significant accuracy loss during the course of the simulation.\\

Measured data are almost perfectly aligned with model predictions, which results in an almost negligible bias and an RMSE which is below 1 $^\circ C$. This excellent result is on par with what already obtained with the very same model on other astronomical sites such as Cerro Paranal and Cerro Armazones in Chile (Lascaux et al. 2013-2014-2015 [\cite{lascauxspie2013,lascauxspie2014,lascaux2015}]), showing the reliability of Meso-Nh in reproducing the base thermodynamic parameters at ground level.\\
\begin{table}[h]
\caption{3$\times$3 contingency table for the absolute temperature during the night, at 55.5~m a.g.l. at LBT, for the sample of 22 nights.
We use the Meso-Nh ${\Delta}$X~=~500~m configuration.} 
\label{tab:tempc}
\begin{center}
\begin{tabular}{cc|ccc}
\hline
\multicolumn{2}{c}{Division by tertiles (climatology)} & \multicolumn{3}{c}{\bf OBSERVATIONS}\\
\multicolumn{2}{c}{MtG - ~55.5~m} & T$<$-1.2 $^{\circ}C$ & -1.2 $^{\circ}C$$<$T$<$2.5 $^{\circ}C$   &  T$>$2.5 $^{\circ}C$ \\
\hline
\multirow{7}{*}{\rotatebox{90}{\bf MODEL}} & & &\\
 & T$<$-1.2 $^{\circ}C$               & 239 & 11 & 0\\
 &      &  &  & \\
 & -1.2 $^{\circ}C$$<$T$<$2.5 $^{\circ}C$ & 14 & 351 & 24\\
 &      &  &  & \\
 & T$>$2.5 $^{\circ}C$             & 0 & 58 & 214\\
 &      &  &  & \\
\hline
\\
\multicolumn{5}{l}{Sample size = 911; PC=88.3\%; EBD=0\%; POD$_1$=94.5\%; POD$_2$=83.6\%; POD$_3$=89.9\%} \\
\end{tabular}
\end{center}
\end{table}
We also report in Table \ref{tab:tempc} the contingency tables obtaind for the temperature parameters with respect to the climatological winter tertiles defined in Table \ref{tab:tertemp}. The model has an impressive PC=88.3\% and a zero EBD, showing almost the same performance in all the tertiles. The results are similar to what obtained in previous works on a much larger statistical sample [\cite{lascaux2015}] of 129 nights, so we expect that the forthcoming final validation study will confirm the present situation.\\

\subsection{Relative humidity}
The relative humidity (RH) with respect to water is the ratio (expressed in percentage) of the actual mixing ratio $w$ to the saturation mixing ratio $w_s$ with respect to water at the same temperature and pressure: $RH = \frac{w}{w_s} \times 100$.\\

\begin{figure}[ht]
\begin{center}
\begin{tabular}{cc}
\includegraphics[height=5cm]{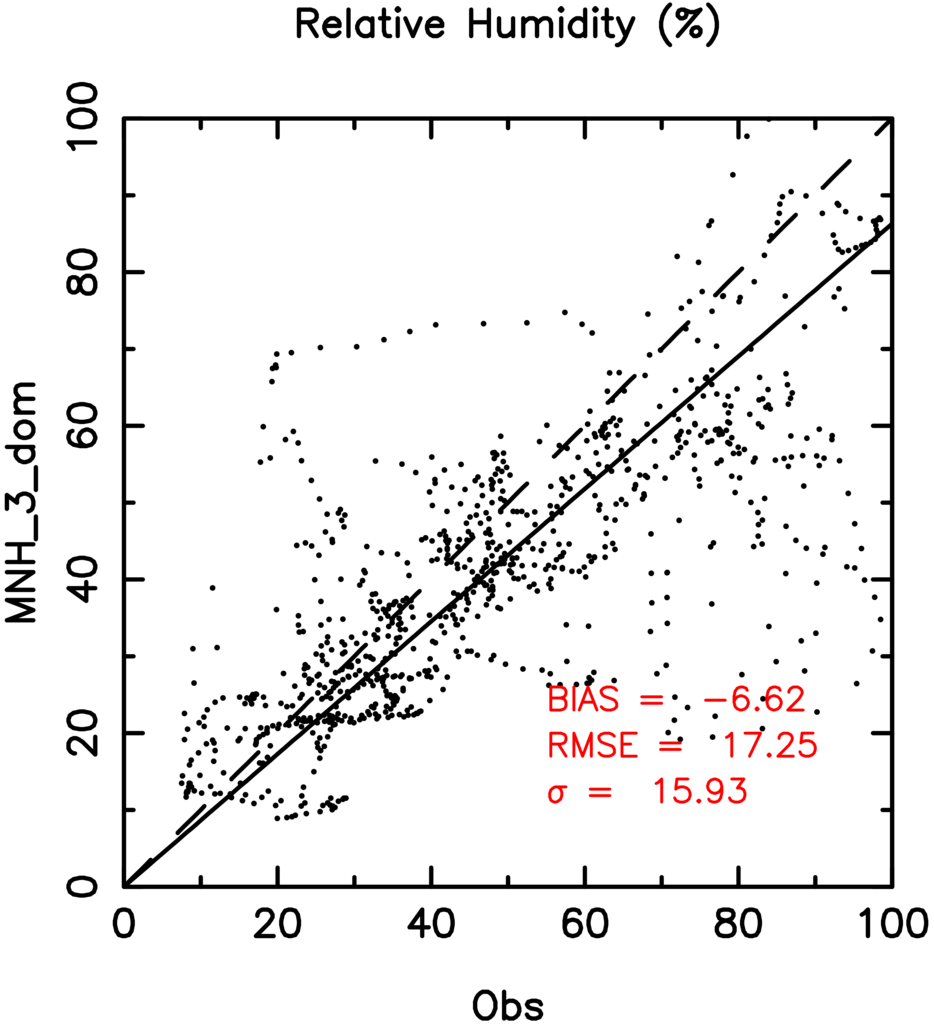} & \includegraphics[height=5cm]{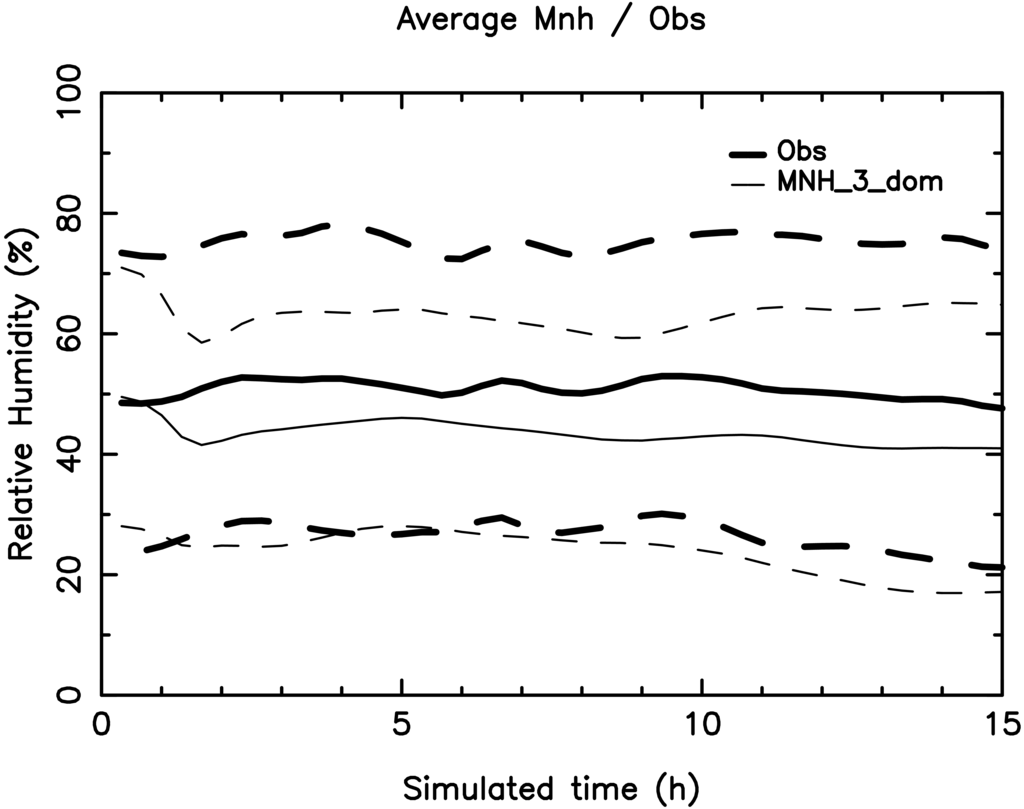} \\
\end{tabular}
\end{center}
\caption[example]{\label{fig:rhreg}Left: scatter plot for relative humidity, comparing model outputs (MNH) and measurements (OBS). The full black line is the regression line passing by the origin, while the dashed line represent the reference diagonal line for unbiased results.\\
Right: time evolution of the average of both model outputs (thin line) and measurements (bold line), from 00:00UT to 15:00UT.} 
\end{figure}

For this parameter, similar to what observed in the case of temperature, we used the model output at the spatial resolution of 500~m. In Fig.\ref{fig:rhreg} (left side) the scatter plot is shown. It is possible to observe that the overall result is good, with measurements and simulated values globally allgned on the diagonal of the plot. The bias is low ($\sim$ -6.6 \%) and the RMS is below 20\%, which is an improvement over previous results obtained at Paranal and Armazones [\cite{lascaux2015}].\\

We notice however that when the measured RH is above $\sim 70\%$, the model systematically tends to underestimate the values, resulting in a poor performance for high values of RH. In other words there is a sort of saturation effect. From observing the time evolution of the average values of both Meso-Nh forecasts and measurements, in Fig.\ref{fig:rhreg}, we notice that this effects starts at the spin-off of the simulation (within the first simulated hour). We also note that the correlation between observed and simulated RH related to a different model level (but close to the level 4 correspondent to the height of the sensor),  provides an improvement of POD$_{3}$ of almost 10\%. This means we have space to improve model performances in this respect. \\
\begin{table}[h]
\caption{3$\times$3 contingency table for the Relative Humidity during the night, at 55.5~m a.g.l. at LBT, for the sample of 22 nights.
We use the Meso-Nh ${\Delta}$X~=~500~m configuration.}
\label{tab:rhc}
\begin{center}
\begin{tabular}{cc|ccc}
\hline
\multicolumn{2}{c}{Division by tertiles (climatology)} & \multicolumn{3}{c}{\bf OBSERVATIONS}\\
\multicolumn{2}{c}{MtG - ~55.5~m} & RH$<$28.4\% & 28.4\%$<$RH$<$71.7\%   &  RH$>$71.7\% \\
\hline
\multirow{7}{*}{\rotatebox{90}{\bf MODEL}} & & &\\
 & RH$<$28.4\%               & 173 & 105 & 11\\
 &      &  &  & \\
 & 28.4\%$<$RH$<$71.7\% & 51 & 377 & 126\\
 &      &  &  & \\
 & RH$>$71.7\%             & 0 & 8 & 60\\
 &      &  &  & \\
\hline
\\
\multicolumn{5}{l}{Sample size = 911; PC=67.0\%; EBD=1.2\%; POD$_1$=77.2\%; POD$_2$=76.9\%; POD$_3$=30.5\%} \\
\end{tabular}
\end{center}
\end{table}

The above situation emerges clearly from the contingency table in Table \ref{tab:rhc}. While the PODs in the first and second tertiles are all excellent ($\sim 77\%$), with a small $EBD=1.2\%$, the POD in the last tertile is similar to the random reference case (33\%). This preliminary study allowed us to individuate this problem, that will be dealt with during the final planned validation study.\\

\subsection{Wind speed}
In the wind speed case, as already observed on other test sites [\cite{masciadri2013,lascaux2013,lascaux2015}], we had to use a spatial resolution $\Delta$X ~=~100~m in order to correctly resolve the strong winds ($\geq 10$~m/s). It is observed [\cite{lascaux2015}] that in the case of lower resolutions ($\Delta$X ~=~500~m) the wind speed is correctly reproduced until it reaches values around $10$~m/s, while above this threshold point the model starts to consistently underestimate the wind speed. We proved in the past that this behaviour is dependent on the model horizontal resolution and it can be almost completely corrected using a higher horizontal resolution.\\
\begin{figure}[ht]
\begin{center}
\includegraphics[height=5cm]{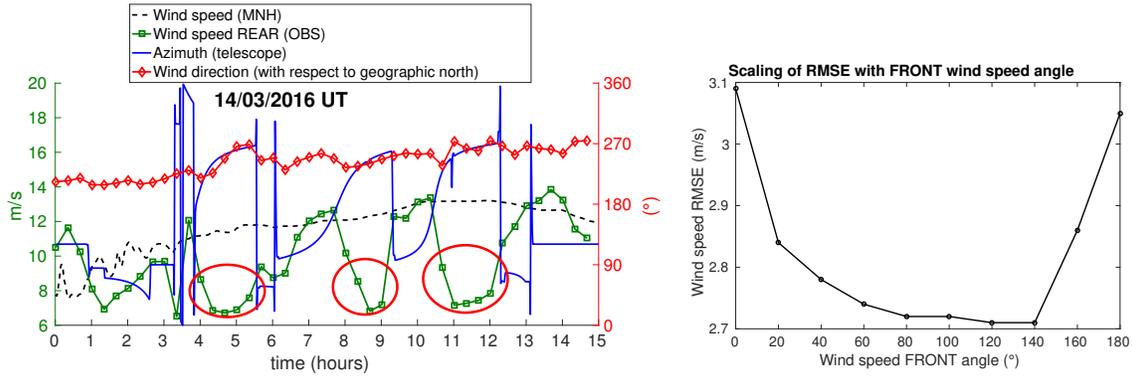}
\end{center}
\caption[example]{\label{fig:az}Left: time evolution of different parameters along the test night of 14/03/2016 UT. The full blue line represents the telescope azimuth (right y-axis showing the angles), the full red line with diamond dots represents the incoming observed wind direction, the full green line with square dots represents the wind speed measured on the REAR anemometer, while the dashed black line represents the model output. With red circles the events in which the wind speed value drops drastically are highlighted, These events coincide with the telescope azimuth facing the incoming wind direction.\\
Right: we display the variation of the RMSE between model outputs and wind speed measurements, with respect to the total angle (centered on the front of the dome) from which we measure the incoming wind speed from the FRONT anemometer. Example: 60$^{\circ}$ means $\pm$ 30$^{\circ}$ with respect to the direction of the telescope in front to the incoming wind. When the wind comes from the complementary angle, we consider measurements from the REAR anemometer.
}
\end{figure} 

In our initial analysis we used as a reference the wind speed measurements coming from the REAR anemometer (which is positioned higher and should be less susceptible to dome-related effects). By comparing these measurements them with those rom the FRONT anemometer, we discovered that they can differ sometime or even 3-4~m/s. 
If we compare measurements from the REAR anemometer with the model outputs, we notice that, when the telescope rotates to face the incoming wind (i.e wind direction close to the telescope azimuth), the observed wind speed  tend to drastically drop, while from other directions it is generally more consistent with the model forecast and more stable with the telescope rotation. This let us think that the wind speed retrieved from the REAR anemometer is not reliable when the telescope is in front to the incoming wind. An example of this effect is portrayed in Fig.\ref{fig:az}, where we selected the case of night (14/03/2016). The red circles in the figure highlight the unrealistic sudden drops in the wind speed, which coincide with the telescope azimuth being approximately coincident with the incoming wind direction. When the telescope rotates away from the incoming wind then the observed wind speed tend to agree with the model prediction. The analysis tend to give the same results if we compare the FRONT anemometer values with respect to wind speed coming from the opposite direction with respect to the dome. A possible explanation for this effect is a drag force produced by the nearby dome surface, being less evident on the REAR anemometer which is positioned higher.\\
For the above reason, we decided to use wind speed from the REAR anemometer when the measured wind direction comes from angles in the [30$^\circ$, 330$^\circ$] range with respect to the telescope azimuth, and use the wind speed from the FRONT one when the wind direction is between [330$^\circ$, 30$^\circ$] with respect to the same axis. We selected those values for the angles based on an analysis reported in Fig.\ref{fig:az} (right side). The RMSE value of the comparison between measurements and model outputs tend to be minimal if we consider the FRONT anemometer for wind speed values coming from an angle which ranges from 60$^\circ$ to 140$^\circ$, centered on the front of the dome, and the REAR measurements for the complementary angles.\\
\begin{figure}[ht]
\begin{center}
\begin{tabular}{cc}
\includegraphics[height=5cm]{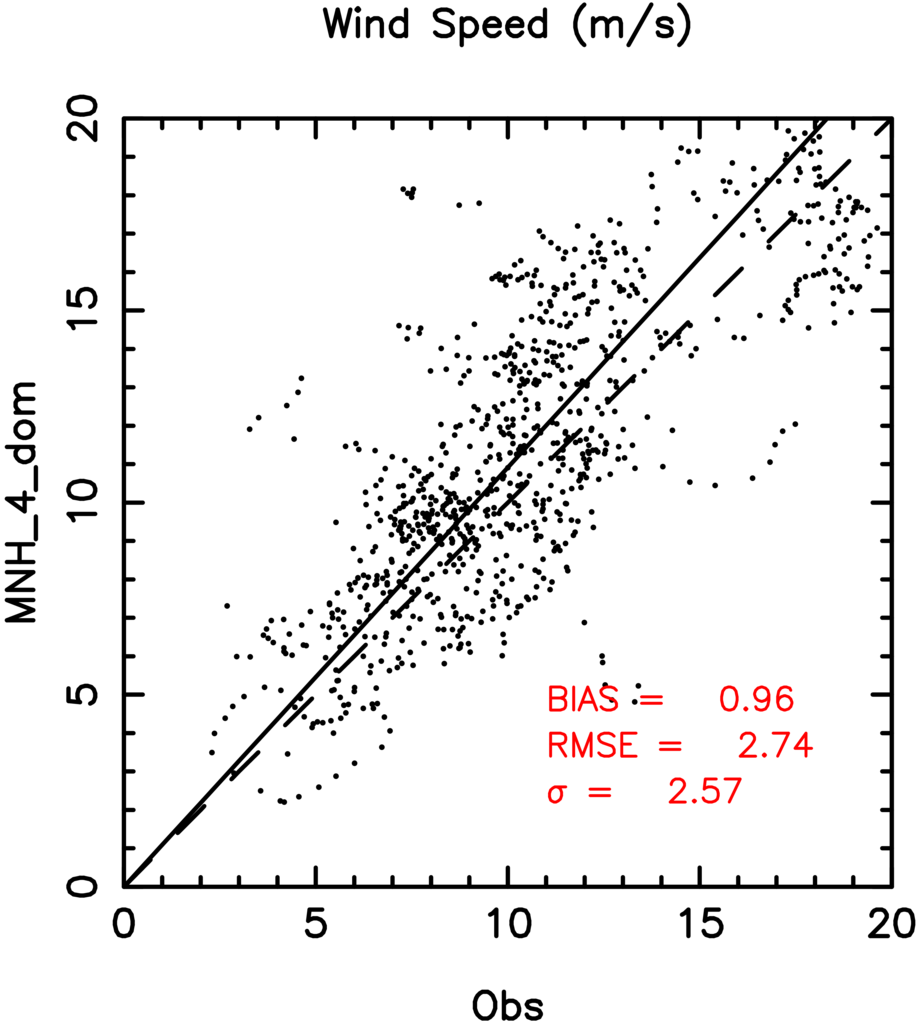} & \includegraphics[height=5cm]{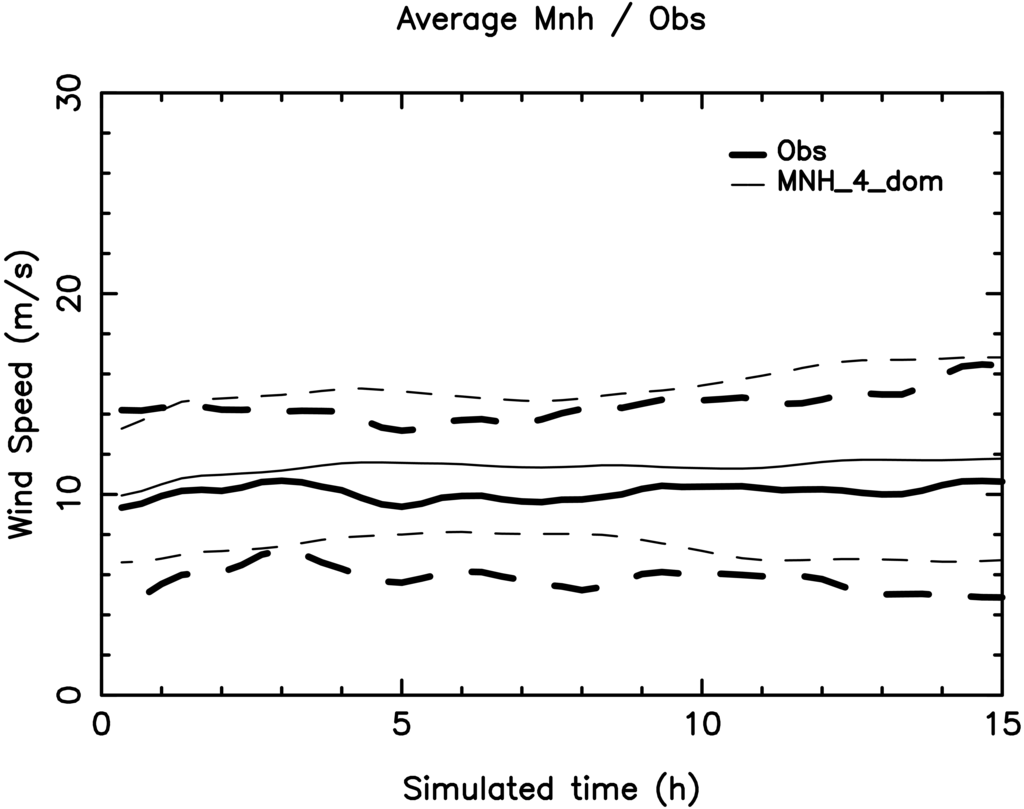} \\
\end{tabular}
\end{center}
\caption[example]{\label{fig:wsreg}Left: scatter plot for wind speed, comparing model outputs (MNH) and measurements (OBS). The full black line is the regression line passing by the origin, while the dashed line represent the reference diagonal line for unbiased results.\\
Right: time evolution of the average of both model outputs (thin line) and measurements (bold line), from 00:00UT to 15:00UT.}
\end{figure} 

Figure \ref{fig:wsreg} shows the scatter plot for the wind speed, where we can observe that measurements and model are correctly aligned independently of the value of the wind speed, with no significant performance loss along the course of the simulation (Fig.\ref{fig:wsreg} right side). The values obtained for the total bias and RMSE are totally consistent with the ones found at Paranal or Cerro Armazones [\cite{lascaux2015}] confirming the validity of the model in resolving the wind speed at low heights.\\
\begin{table}[h]
\caption{3$\times$3 contingency table for the wind speed during the night, at 58~m a.g.l. at LBT, for the sample of 22 nights. We use the Meso-Nh ${\Delta}$X~=~100~m configuration.}
\label{tab:ws}
\begin{center}
\begin{tabular}{cc|ccc}
\hline
\multicolumn{2}{c}{Division by tertiles (climatology)} & \multicolumn{3}{c}{\bf OBSERVATIONS}\\
\multicolumn{2}{c}{MtG - ~58~m} & WS$<$5.9~m/s & 5.9~m/s$<$WS$<$9.7~m/s   &  WS$>$9.7~m/s \\
\hline
\multirow{7}{*}{\rotatebox{90}{\bf MODEL}} & & &\\
 & WS$<$5.9~m/s               & 35 & 19 & 5\\
 &      &  &  & \\
 & 5.9~m/s$<$WS$<$9.7~m/s & 35 & 183 & 58\\
 &      &  &  & \\
 & WS$>$9.7~m/s             & 7 & 135 & 435\\
 &      &  &  & \\
\hline
\\
\multicolumn{5}{l}{Sample size = 911; PC=71.6\%; EBD=1.3\%; POD$_1$=44.7\%; POD$_2$=54.3\%; POD$_3$=87.3\%} \\
\end{tabular}
\end{center}
\end{table}

When analysing the contingency table (Table \ref{tab:ws}), we see that POD$_{i}$ are excellent especially in the third tertile (POD$_{3}$=$\sim 87\%$), which is the most interesting case since the conditions of strong wind speed are the most critical from an astronomical point of view. They are indeed among the principal causes of  vibrations propagated through the telescope mount. The $EBD=1.3\%$ is very small, and the lower POD in the first tertile is of little importance since it regards wind speed values which are almost negligible for the telescope operation.\\

\subsection{Wind direction}
\label{wdchapt}
The wind direction convention for angles used in this paper, unless otherwise specified, is with respect to the geographical north, counting clockwise ($0^\circ$=North, $90^\circ$=East, etc...). In order to compare the model results to the observed values, we discarded the wind direction values which are relative to low wind speeds $<3$~m/s. This was also done in previous works [\cite{lascaux2015}], for two reasons. On one side when the wind speed is low the wind direction tend to have a large dispersion due to random oscillations, while on the other it is less interesting to measure wind direction when there is almost no wind at all. This criterion allowed us to discard 7 measurements form the whole statistical sample, leaving us with 904 valid wind direction measurements over the 22 nights sample in Table \ref{tab:dates}. In the case of the present preliminary validation study the application of this criterion is negligible, however we do expect to have a sensible effect using this methodology on the final results of the validation study on a larger sample.\\
\begin{figure}[ht]
\begin{center}
\begin{tabular}{cc}
\includegraphics[height=5cm]{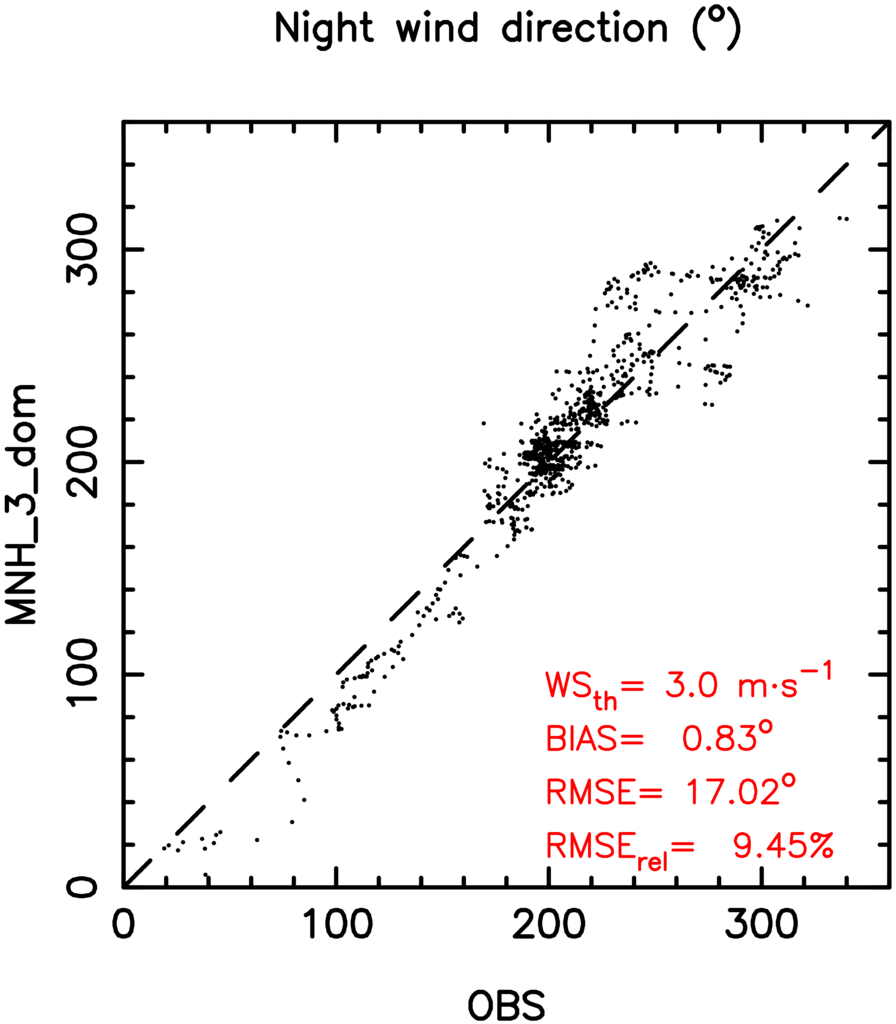} & \includegraphics[height=5cm]{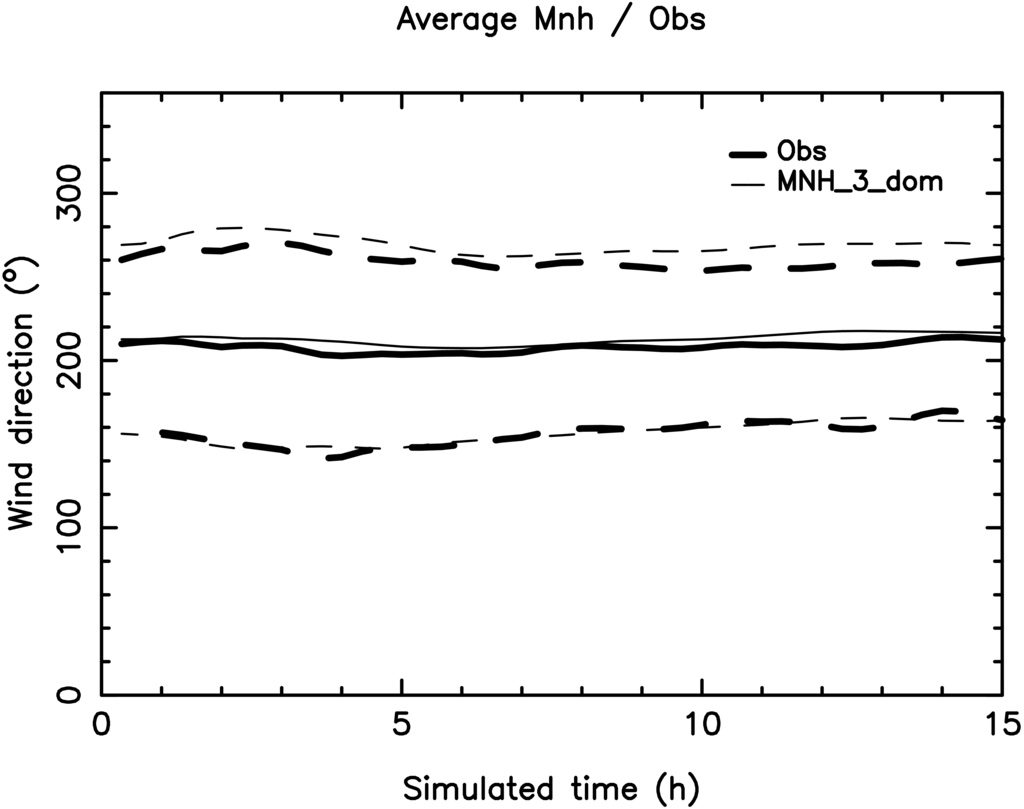} \\
\end{tabular}
\end{center}
\caption[example]{\label{fig:wdreg}Left: scatter plot for wind direction, comparing model outputs (MNH) and measurements (OBS). The full black line is the regression line passing by the origin, while the dashed line represent the reference diagonal line for unbiased results.\\
Right: time evolution of the average of both model outputs (thin line) and measurements (bold line), from 00:00UT to 15:00UT.}
\end{figure}

As introduced in Lascaux et al. 2015, we calculated another statistical quantity $RMSE_{rel} = \frac{RMSE}{180^\circ} \times 100$ i.e. the relative error of the RMSE calculated obviously with respect the the worst possible prediction (reversed by $180^\circ$).\\
\begin{table}[h]
\caption{4$\times$4 contingency table for the wind direction during the night, at 58m a.g.l. at LBT, for the sample of 22 nights.
We use the Meso-Nh ${\Delta}$X~=~500~m configuration.}
\label{tab:wdcon}
\begin{center}
\begin{tabular}{cc|cccccc}
\multicolumn{2}{c}{ } & \multicolumn{4}{c}{\bf OBSERVATIONS}\\
\multicolumn{2}{c}{MtG - ~58~m} & North & East & South & West \\
\hline
\multirow{9}{*}{\rotatebox{90}{\bf MODEL}} & & &\\
 & North & 43 & 4 & 0 & 1 \\
 &   & & & & \\
 & East & 0 & 57 & 14 & 0 \\
 & & & & & \\
 & South & 0 & 0 & 463 & 12 \\
 &   & & & & \\
 & West & 8 & 0 & 75 & 227 \\
 &   & & & & \\
\hline
\\
\multicolumn{5}{l}{Sample size = 911; PC=87.4\%; EBD=0\%} \\
\multicolumn{5}{l}{POD$_N$=84.3\%; POD$_E$=93.4\%} \\
\multicolumn{5}{l}{POD$_S$=83.9\%; POD$_W$=94.6\%} \\
\end{tabular}
\end{center}
\end{table}

Model predictions for the wind direction near ground, as it can be seen from the scatter plot of Fig.\ref{fig:wdreg}, show a consistently increased accuracy with respect to the results obtained in previous works [\cite{lascaux2015}]. While on Cerro Paranal we were able to detect the wind direction with an $RMSE\simeq35^\circ$, in this case the RMSE is $\sim$ 17$^\circ$. This corresponds to an RMSE$_{rel}$$\simeq$ 9.5\%, which definitely allows the telescope operator to successfully identify the incoming wind direction within the operational margins. The temporal evolution of the average values also shows how forecasts and measurements tend to be correctly aligned along the course of the whole simulation. While the validation study is ongoing and this results are valid only for a small preliminary sample, they are, nonetheless, encouraging.\\
The extremely positive performance of the model is confirmed by the contingency table (Table \ref{tab:wdcon}). In this case we selected all wind direction measurements, even below 3~m/s, and we divided the whole angle in 4 quadrant (North, East, South and West), using a 4$\times$4 table. The global PC scores an almost perfect $87\%$ with a zero EBD and the same level of performance in all the PODs relative to the different quadrants.\\

\section{CONCLUSIONS}
\label{sec:concl} 
In this paper we presented the preliminary results of the ongoing validation study on the operational forecast system being developed for LBT, as part of the ALTA project, performed on a limited 22 nights test sample. We performed this test on the atmospherical parameters measured near the ground level, using as a reference the weather instruments of the LBT dome, in the context of the first commissioning for the atmospheric forecasts in June 2016. The results we obtained are extraordinarily good and show excellent performance in reconstructing all the tested parameters. Temperature is predicted with an excellent degree of accuracy, with an RMSE~$=0.98 ^\circ C$ and a PC~=~88.3\% with respect to the winter distribution of measurements. Wind speed is reconstructed with an RMSE~=~2.7~m/s and a PC~=~71.6\%, which raises to a POD~$=87.3\%$ in the case of strong winds ($>10$~m/s). The result we obtained for the wind direction is incredibly positive, with an RMSE~$=17^\circ$, RMSE$_{rel}$$\simeq$ 9.5\% and a PC~=~87.4\%, showing a major increase of accuracy with respect to the previous studies conducted on ESO telescope sites. The result for the relative humidity is good, with a global RMSE~=~17.3\%, however we still have problems in correctly detecting high values of the parameter, leading to a unsatisfactory prediction when the value is higher than $\sim 70\%$. This problem will be addressed before the forthcoming final validation study that will be performed on a larger statistical sample of nights.

\acknowledgments  
ALTA project is funded by the Large Binocular Telescope Corporation. Measurements of surface parameters are provided by the LBT telemetry and annexed instrumentation. The authors thanks Christian Veillet, Director of the Large Binocular Telescope, for his continued and valuable support given to this research activity. Authors also thanks the LBTO staff for their technical support and collaboration. Part of the numerical simulations have been run on the HPCF cluster of the European Center for Medium Range Weather Forecasts (ECMWF) using resources from the Project SPITFOT.



\end{document}